# Coherent spin-wave transport in an antiferromagnet

J.R. Hortensius[1‡*], D. Afanasiev[1‡†], M. Matthiesen[1], R. Leenders[2], R. Citro[3], A.V. Kimel[4], R.V. Mikhaylovskiy[2], B.A. Ivanov[5], and A.D. Caviglia[1§]

[1]*Kavli Institute of Nanoscience, Delft University of Technology, P.O. Box 5046, 2600 GA Delft, The Netherlands.*
[2]*Department of Physics, Lancaster University, Bailrigg, Lancaster LA1 4YW, United Kingdom*
[3]*Dipartimento di Fisica "E.R. Caianiello", Università di Salerno and Spin-CNR, I-84084 Fisciano (Sa), Italy*
[4] *Institute for Molecules and Materials, Radboud University Nijmegen, 6525 AJ Nijmegen, The Netherlands.*
[5] *Institute of Magnetism, National Academy of Sciences and Ministry of Education and Science, 03142 Kyiv, Ukraine.*

‡These authors contributed equally to this work
* j.r.hortensius@tudelft.nl
† dmytro.afanasiev@physik.uni-regensburg.de
§ a.caviglia@tudelft.nl

**Magnonics is a research field complementary to spintronics, in which the quanta of spin waves (magnons) replace electrons as information carriers, promising less energy dissipation[1-3]. The development of ultrafast nanoscale magnonic logic circuits calls for new tools and materials to generate coherent spin waves with frequencies as high, and wavelengths as short, as possible[4,5]. Antiferromagnets can host spin waves at THz frequencies and are therefore seen as a future platform for the fastest and the least dissipative transfer of information[6-11]. However, the generation of short-wavelength coherent propagating magnons in antiferromagnets has so far remained elusive. Here we report the efficient emission and detection of a nanometer-scale wavepacket of coherent propagating magnons in antiferromagnetic $DyFeO_3$ using ultrashort pulses of light. The subwavelength nanoscale confinement of the laser field due to large absorption creates a strongly non-uniform spin excitation profile, thereby enabling the propagation of a broadband continuum of coherent THz spin waves. The wavepacket features magnons with detected wavelengths down to 125 nm and supersonic velocities up to 13 km/s that propagate over macroscopic distances. The long-sought source of coherent short-wavelength spin carriers demonstrated here opens up new prospects for THz antiferromagnetic magnonics and coherence mediated logic devices at THz frequencies.**

Antiferromagnetic insulators (AFMs) are prime candidates to replace ferromagnets (FMs) as active media in the quest towards high-speed spin transport and large spectral bandwidth operation[6-8]. Integration of AFMs in future wave-based technologies[3] crucially requires realization of coherent (ballistic) transport of antiferromagnetic spin waves over large distances[5]. Non-uniform spin-wave modes with short wavelengths ($\lambda \lesssim 100$ nm) are particularly important: they can operate at THz clock rates, exhibit high propagation velocities and enable the miniaturization of devices down to the nanoscale. Phase-coherent ballistic spin transport in AFMs is also interesting from a fundamental point of view, as it is predicted to be a prerequisite for the occurrence of exotic phenomena such as magnetic solitons[12], Bose-Einstein condensates[13,14] and spin-superfluidity[15-17]. These prospects call for efficient methods for the excitation, manipulation, and detection of short-wavelength coherent antiferromagnetic magnons.

Conventional methods of linear spin-wave excitation use spatially varying oscillating magnetic fields. The high-frequency THz resonances inherent to antiferromagnetic dynamics make traditional field sources based on microstrip lines or coplanar waveguides impractical to be used in antiferromagnetic media. As a result, recent demonstrations of magnon-mediated spin transport in antiferromagnets were represented either by diffusive propagation of incoherent magnons[9-11] or by evanescent spin-wave modes[18], and the generation of coherent propagating spin waves in an antiferromagnet has yet to be experimentally realized.

Ultrashort pulses of light have been routinely used to generate and to control large-amplitude THz spin precession[19-21] in antiferromagnets. The small photon momentum, however, poses a problem: it gives rise to a large momentum mismatch with short-wavelength spin waves. Consequently, optical techniques have so far been restricted to the generation of $k = 0$ uniform antiferromagnetic magnons, for which group velocities are (near-)zero and no spatial transport of energy and angular momentum takes place. Here we overcome this problem and present an all-optical method to excite and detect a broadband wavepacket of coherent short-wavelength propagating magnons in an insulating antiferromagnet using ultrashort pulses of light. Excitation of intense charge-transfer electronic transitions in the prototypical antiferromagnet $DyFeO_3$ provides strong confinement of the optical field, which creates an exponential profile of non-uniformly deflected spins near the sample surface. This magnetic non-uniformity extends over the nanoscale penetration depth of the light and serves as a source of short-wavelength coherent spin waves propagating into the sample bulk, as illustrated in Figure 1a. Using $k$-selective magneto-optical detection we map out spectral components of the magnon wavepacket and reveal magnon modes with nanoscale wavelengths and supersonic group velocities propagating over macroscopic distances.

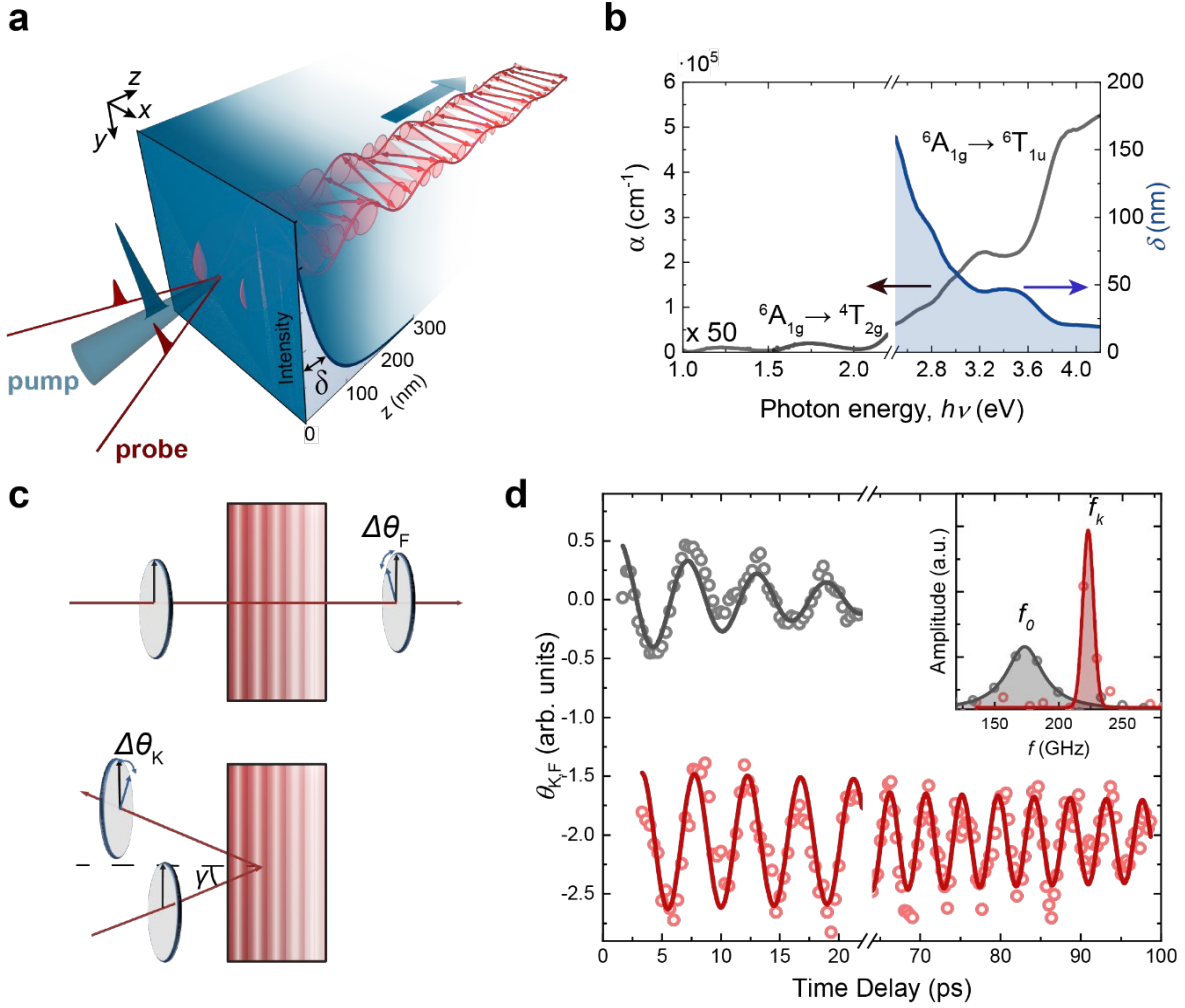

**Figure 1: All-optical generation and detection of coherent antiferromagnetic spin waves.** (a) Schematic illustration of the generation and detection of propagating antiferromagnetic spin waves after excitation of strongly absorbing electronic transitions. The optical penetration depth $\delta$ of the light defines the excited region. (b) Absorption coefficient (left axis) and penetration depth (right axis) for $DyFeO_3$ as function of photon energy (see methods). (c) Schematics for the optical detection mechanisms of spin waves in transmission (top) and reflection (bottom) geometries, measuring transient changes in the Faraday rotation ($\theta_F$) and Kerr rotation ($\theta_K$) respectively. $\gamma$: angle of incidence. (d) Time-resolved measurements of the polarization rotation of a near-infrared probe pulse after excitation with pump pulses with a photon energy of 3.1 eV in the detection geometries shown in (c). The thick solid lines are exponentially damped sine fits. Inset: Fourier spectra of the oscillations with Lorentzian fits (thick solid lines), with central frequencies $f_0$ (transmission geometry) and $f_k$ (reflection geometry). a.u.: arbitrary units.

Dysprosium orthoferrite ($DyFeO_3$) is a charge-transfer antiferromagnetic insulator with the Néel temperature $T_N$ = 645 K, exhibiting one of the strongest observed interactions between spins and laser pulses[19,22]. The optical spectrum of $DyFeO_3$ is dominated by a set of intense electronic O-Fe ($2p$-$3d$) charge-transfer (CT) transitions. The absorption due to these transitions sets in above 2 eV, and promptly brings the absorption coefficient to values as high as $5 \times 10^5$

cm$^{-1}$ (see Fig. 1b)[23]. Such strong absorption enables confinement of light down to penetration depths ($\delta$) of less than 50 nm.

In our experiments we study a 60 μm thick slab of z-cut DyFeO$_3$. The sample is excited with 100 fs pump pulses which have photon energy tunable in the spectral range of 1.5 - 3.1 eV, covering the lowest energy $^6A_{1g} \rightarrow {}^6T_{1u}$ charge-transfer electronic transition[23]. We use time-delayed probe pulses at various photon energies below the fundamental absorption gap ($h\nu < 2$ eV) to detect the photo-induced magnetic dynamics, see Extended Data Fig. 1. The probing is simultaneously performed in two complementary experimental geometries: a transmission geometry using the Faraday effect and a reflection geometry, using the magneto-optical Kerr effect (MOKE) (see Fig. 1c). In both geometries, the pump-induced rotation of the probe polarization plane, originating from the Faraday effect ($\theta_F$) or the MOKE ($\theta_K$), is tracked as a function of the pump-probe time delay. Note that while the Faraday transmission geometry is routinely used in pump-probe experiments for detecting uniform ($k = 0$) spin precession in antiferromagnets[19], the reflection geometry has been shown to enable detection of finite-$k$ coherent excitations such as propagating acoustic wavefronts[24]. As shown below we demonstrate that the reflection geometry can be also used to probe the dynamics of short-wavelength propagating coherent spin waves.

Following the optical pumping in the regime of strong absorption ($h\nu = 3.1$ eV, $\delta = 50$ nm) the dynamics of the probe polarization reveal high-frequency oscillations in the hundreds of GHz range (see Fig. 1d). The frequencies $f_0$ and $f_k$ of the oscillations observed in the transmission and reflection geometry respectively, are substantially different: $f_k > f_0$ (see inset Fig. 1d). Notably, the decay time of the oscillations also differ by nearly an order of magnitude.

To identify the origin of the oscillations, we track their central frequency as a function of temperature. The antiferromagnetic state in DyFeO$_3$ adopts two distinct spin arrangements, sharply separated by a first-order phase transition at the so-called Morin temperature $T_M \simeq 50$ K[25]. At $T < T_M$, the antiparallel iron spins are oriented along the y-axis and arranged in a compensated collinear AFM pattern; while above $T_M$, they experience a spin-reorientation towards the x-axis accompanied by the mutual canting and stabilization of a canted AFM phase (see Figure 2a). The experimentally acquired temperature dependence of the oscillation frequency exhibits a characteristic cusp-like softening with a minimum at $T_M$ (see Fig. 2b and Extended Data Fig. 2). This frequency softening is an unambiguous hallmark of the so-called quasi-antiferromagnetic ($q$-AFM) magnon branch in DyFeO$_3$ and is caused by strong temperature variations of the magneto-crystalline anisotropy in the vicinity of the spin-reorientation phase transition[26]. Indeed, the frequencies $f_0$ of the oscillations observed in the transmission geometry perfectly match values reported in literature for the zone-centre ($k = 0$) $q$-AFM magnon[26].

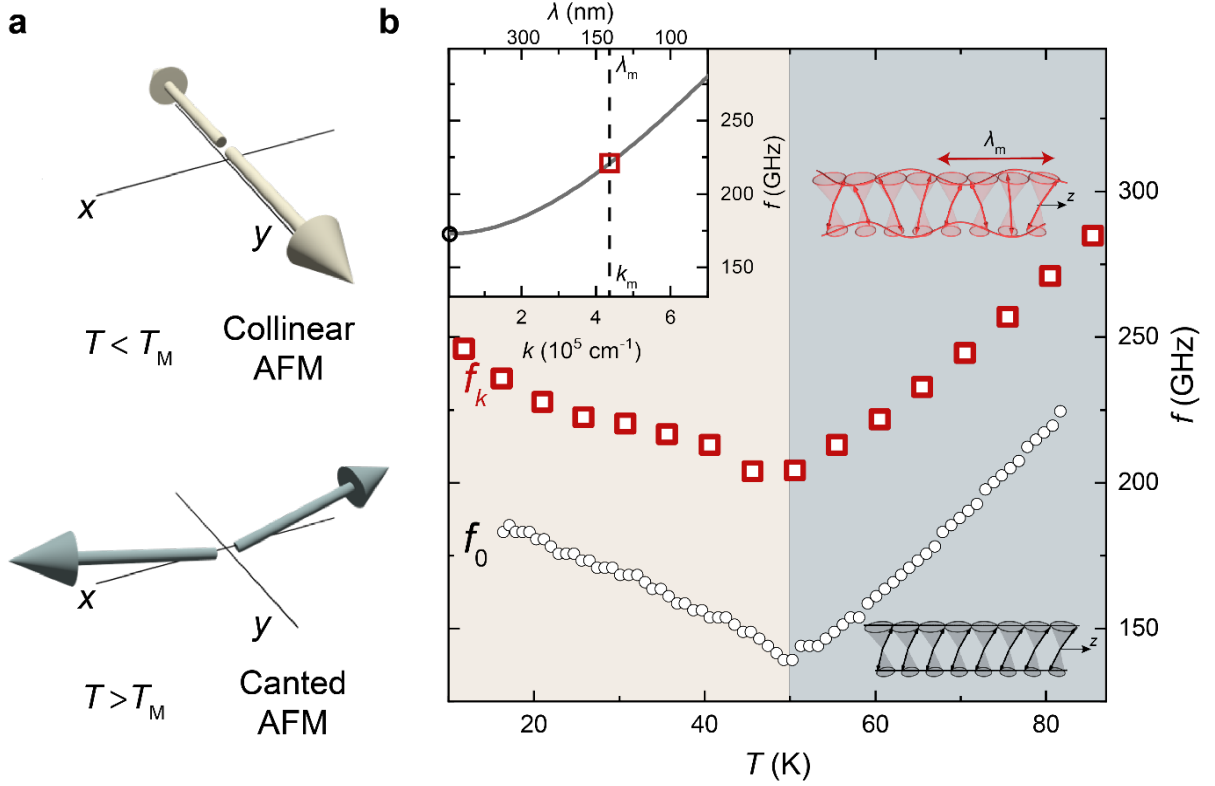

**Figure 2: Generation and detection of coherent finite-$k$ antiferromagnetic spin waves.** (**a**) The spin orientation in the collinear AFM (top panel) and canted AFM phase (bottom panel). (**b**) Temperature ($T$) dependence of the central frequency of the oscillatory dynamics as measured in the reflection geometry (red square markers) after excitation with strongly absorbed pump pulses ($h\nu = 3.1$ eV, $\delta = 50$ nm) compared with the $k = 0$ magnon $T$-dependence, as measured in standard transmission geometry (black circle markers). Left top inset: The antiferromagnetic magnon dispersion of DyFeO$_3$ and the wavenumbers of the magnons observed in the different experimental geometries. Right insets: Schematic illustration of the spin wave corresponding to the oscillatory dynamics at the different frequencies.

To explain the physical origin of the oscillation at frequency $f_k$ seen in the MOKE experiment, we refer to the dispersion relation for magnon modes in antiferromagnets. In both magnetic phases, below and above $T_M$, the magnon spectrum $\omega(k)$ in DyFeO$_3$, is given by[27]:

$$\omega_k = \sqrt{\omega_0^2 + (v_0 k)^2}, \quad (1)$$

where $v_0 \approx 20$ km/s is the limiting group velocity of the spin waves[27]. This dispersion relation is schematically shown as an inset to Fig. 2b. At small wavenumbers $kv_0 \ll \omega_0$, it has a quadratic form due to the magnon gap $\omega_0 = 2\pi f_0$, arising from magneto-crystalline anisotropy. At larger wavenumbers ($v_0 k \gg \omega_0$), the dispersion relation becomes dominated by the exchange interaction (exchange regime), and thus takes a linear form typical for antiferromagnets[27]. The MOKE signal at $f_k$ can be identified as a finite-$k$ magnon on the $q$-AFM branch: it follows the characteristic temperature dependence of the $f_0$ zone-center magnon mode and has a nearly temperature-independent blueshift. The detection geometry implies that the magnon wavevector is perpendicular to the sample surface, and its magnitude can be deduced from Eq. (1) to be $k = 4.2 \times 10^5$ cm$^{-1}$ ($\lambda \approx 140$ nm), as marked on the inset to Figure 2b.

By considering the modulation of the material's dielectric tensor due to the propagating coherent spin wave, the attribution of the $f_k$ oscillation to a finite-$k$ magnon on the $q$-AFM branch can be further supported. A spin wave with a propagation vector along the $z$-axis results in a perturbation of the magnetic order and a corresponding periodic modulation of the off-diagonal components of the dielectric tensor[28]. In this way the spin waves in the medium produce the magneto-optical analogue of a dynamical volume phase grating for the probe light wave. The polarization state of the reflected probe is explained by the Bragg reflection of light from this diffraction grating, an approach similar to the one used in Brillouin light scattering studies on spin waves[29]. As a result, the polarization rotation of the reflected probe beam with wavenumber $k_0$ becomes subject to a Bragg condition,

$$k_\mathrm{m} = 2k_0 n(\lambda_0) \cos\gamma', \qquad (2)$$

where $n(\lambda_0)$ is the optical refractive index of the medium at the probe wavelength $\lambda_0$, $\gamma'$ is the refracted angle of incidence of the probe, and $k_\mathrm{m}$ is the normal projection of the **k**-vector of the magnon to which the probe pulse is sensitive (see Supplementary section S1). Using Eq. (2) we find that a probe pulse at a central wavelength of 680 nm ($n \approx 2.39$)[23] and normal incidence ($\gamma' = 0$) is sensitive to propagating magnons with wavenumber $k \approx 4.2 \times 10^5$ cm$^{-1}$. This independent estimation agrees with the magnon wavenumber retrieved using the measured frequency and known dispersion relation (Eq. (1)).

The generation and detection of finite-$k$ coherent magnons is anticipated to rely strongly on the confinement provided by the optical penetration depth $\delta$, which is highly dispersive near the charge-transfer band. In particular, changing the pump photon energy between 2.4 eV and 3.1 eV provides a variation in the penetration depth between 300 and 50 nm, while the real part of the refractive index (influencing the pump wavelength) changes by only 5% (see Extended Data Fig. 3). Therefore, we expect that by changing the photon energy of the pump pulses, the amplitude of the finite-$k$ magnon will vary strongly. The time-resolved magneto-optical signals obtained in the reflection MOKE geometry, for different photon energies of the pump excitation, are shown in Fig. 3a. The Fourier transforms of the signals (Fig. 3b) show that the spectra are composed of two components, corresponding to the zone-centre and finite-$k$ magnon modes. We observe that with increasing photon energy (i.e. decreasing penetration depth), the amplitude of the finite-$k$ magnon mode substantially increases. The magnon amplitude extracted for the pump excitation at different photon energies, as a function of the corresponding penetration length, is shown in Fig. 3c. The obtained dependence shows that the finite-$k$ magnon is nearly absent for penetration depths larger than 150 nm and grows dramatically for shorter penetration lengths.

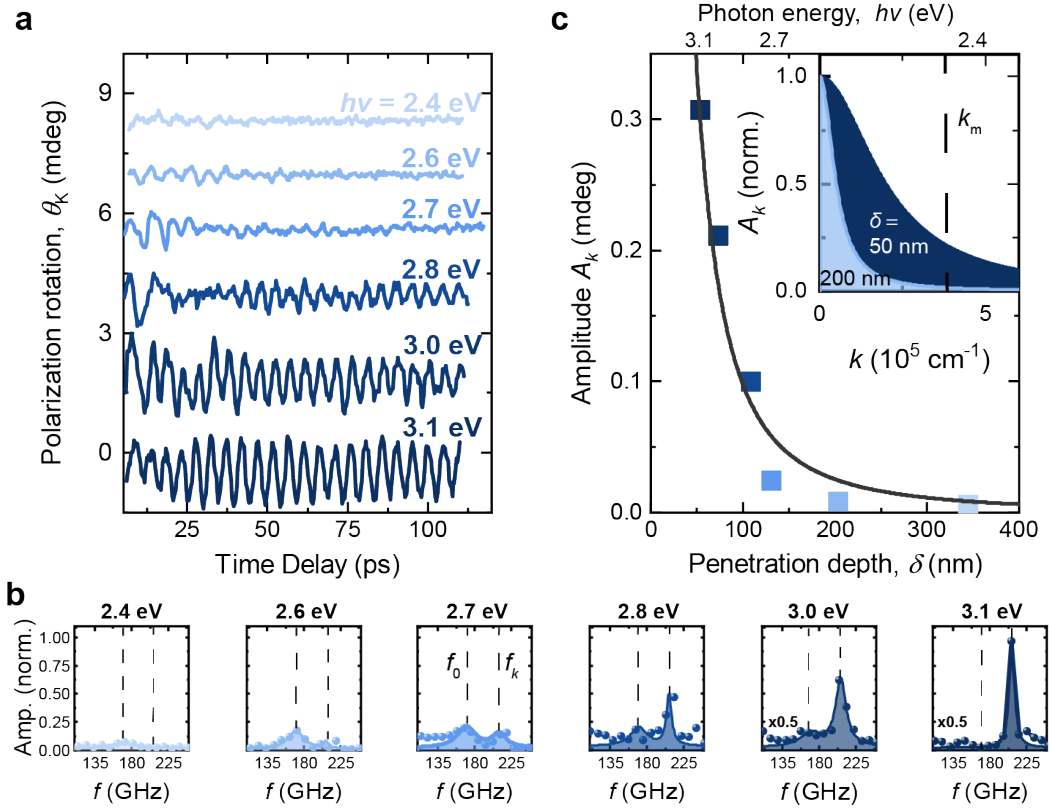

**Figure 3: Confinement of the light as a necessary condition for the generation of high-$k$ spin waves. (a)** Time-resolved signals of the polarization rotation of the near-infrared probe pulse after excitation with pump pulses with increasing photon energy in a reflection geometry. **(b)** Fourier amplitude spectra of the time-resolved signals from panel (a). **(c)** Amplitude of sine fit oscillations corresponding to the AFM propagating spin wave to the data from panel (a) vs penetration depth of the excitation pulse (color markers correspond to traces in panel (a)). The solid line is a fit using $I_0/(1 + (k\delta)^2)$. Inset: Magnon frequency distribution after excitation with pump pulses with $\delta$=50 (dark blue, broad distribution) and 200 nm (light blue, narrow). The probe is sensitive to $k_m = 3.7 \cdot 10^5$ cm$^{-1}$, indicated by the dashed line.

We model this observation using a simple assumption: the ultrashort light pulse promotes a spin excitation that is strongly non-uniform along the direction of incidence $z$. The excitation leads to a nearly instantaneous deflection of spins by an angle $\varphi(z)$ with the spatial distribution following the optical absorption profile according to Beer-Lambert law: $\varphi(z,t=0) = \varphi_0 e^{-z/\delta}$, where $\varphi_0 \sim I_0/\delta$ is proportional to the intensity of the pump pulse $I_0$, and inversely proportional to the light penetration depth $\delta$ (see Supplementary section S2). The strongly non-uniform spin excitation distributes the initial deflection among magnon modes at different $k$-vectors, with the magnon amplitude given by the reciprocal space image $A_k$ of the initial excitation set up by the penetration depth (see inset Fig. 3b). As a result, the finite-$k$ mode is expected to have spectral amplitude $A_k$ (see Supplementary Section S2):

$$A_k \sim \frac{I_0}{1+(k\delta)^2} \quad (3)$$

This expression not only agrees well with the observations of Fig. 3c ($k \approx 3.7\times10^5$ cm$^{-1}$, $\lambda = 170$ nm), where it is plotted as a best fit to the pump intensity $I_0$, but also confirms the intuitive interpretation that a stronger confinement shifts the spectral amplitude of the excited magnon wave packet towards larger $k$.

The excitation of a continuum of coherent antiferromagnetic spin waves forms a broadband magnon wavepacket, in which individual spectral components propagate independently, each adhering to the dispersion relation $\omega_k$. In order to visualize the time evolution of the excited magnon wavepacket, we make use of the linearized sine-Gordon equation for the space- ($z$) and time- ($t$) dependent amplitude of the spin deflections $\varphi(z,t)$[27]. We find that the evolution of the spin dynamics is described by the following simple formula (see Supplementary section S2):

$$\varphi(z,t) = \frac{2}{\pi} \int_{-\infty}^{\infty} dk [A_k \cos(kz) \cos(\omega_k t)] \quad (4)$$

By integrating this equation, under the assumption of the exponential distribution of the time-zero spin deflections and $\delta = 50$ nm, we obtain the dynamics of the magnetic excitation. As shown in Fig. 4a and Extended Data Fig. 4, the strong dispersion promptly smears out the initial exponential profile of the spin excitation simultaneously forming a spin-wave front after ~10 ps that propagates into the bulk. This front is composed by the short wavelength magnons with $k \gtrsim 20\times10^5$ cm$^{-1}$ propagating with the limiting group velocity $v_0$.

Applying the Bragg condition of Eq. (2), we can experimentally map out the spectral components of the excited broadband magnon wave packet, as well as determine the group velocity and propagation length of individual magnon modes. First, we vary the incidence angle $\gamma$ of the probe pulse, (inset Fig. 4b) and find that the central frequency of the oscillations is reduced upon increasing $\gamma'$ (Fig. 4b), in perfect agreement with Eq. (2) and the magnon dispersion of Eq. (1). Next, upon decreasing the probe wavelength, we observe a systematic increase in the magnon frequency, as shown in Fig. 4c, once again in accordance with the Bragg condition. To summarize our observations, we plot the extracted central frequencies as a function of the corresponding wavenumbers $k_m$ (see Fig. 4d). These points are fit to the dispersion relation given by Eq. (1), with the spin-wave gap $\omega_0 = 2\pi f_0$ defined by the experimentally obtained zone-center magnon frequency. The best fit of the experimental data yields a limiting group velocity $v_0 = 19.7 \pm 0.1$ km/s. This value stands in excellent agreement with the limiting group velocity of 20 km/s extracted from the speed limit of the magnetic domain walls in DyFeO$_3$, as reported in Ref. [27]. Using the extracted value of the limiting group velocity we evaluate the group velocities $v_g = \frac{\partial \omega_k}{\partial k}|_{k=k_m}$ of the optically detected magnons given by $v_g = v_0^2 \frac{k_m}{\omega_k}$. These values, shown in Fig. 4e, indicate that the shortest-wavelength components of the magnon wavepacket detected in our experiment ($\lambda = 125$ nm) propagate at a supersonic ($v_s = 6$ km/s, see Supplementary Section S4) velocity of nearly 13 km/s. We note that magnons with these wavelengths already approach the exchange wave regime characterized by the limiting group velocity $v_0$. This remarkable feature, inherent to antiferromagnets, stands in sharp contrast with the situation in ferromagnets, where the quadratic dispersion relation dictates that the exchange value of the group velocity is reached only for magnons with $\lambda \lesssim 10$ nm. Although the shortest magnon wavelength detected in our experiments is 125 nm, magnons at even shorter wavelengths, down to the penetration depth limit of 50 nm, are anticipated, and could be detected using probe pulses at higher photon energies or other means. Using the extracted lifetime of the oscillations $\tau = 85$ ps (see Supplementary Section S3), we estimate the coherence length $l_c$ of the spin-wave transport $l_c = v_g \tau = 1.1$ μm. We note that this length, dramatically enhanced as compared to metallic antiferromagnets[8,30], also agrees with studies of diffusive spin transport in other insulating antiferromagnets[9,31]. One can anticipate even longer propagation lengths for the coherent (ballistic) regime reported here: our estimate of the coherence length is only a lower limit, as the propagating spin wave is likely to escape from the region that is probed by the reflected probe light. These striking observations make antiferromagnetic insulators such as DyFeO$_3$ a promising platform for the realization of high-speed wave-based magnonic devices.

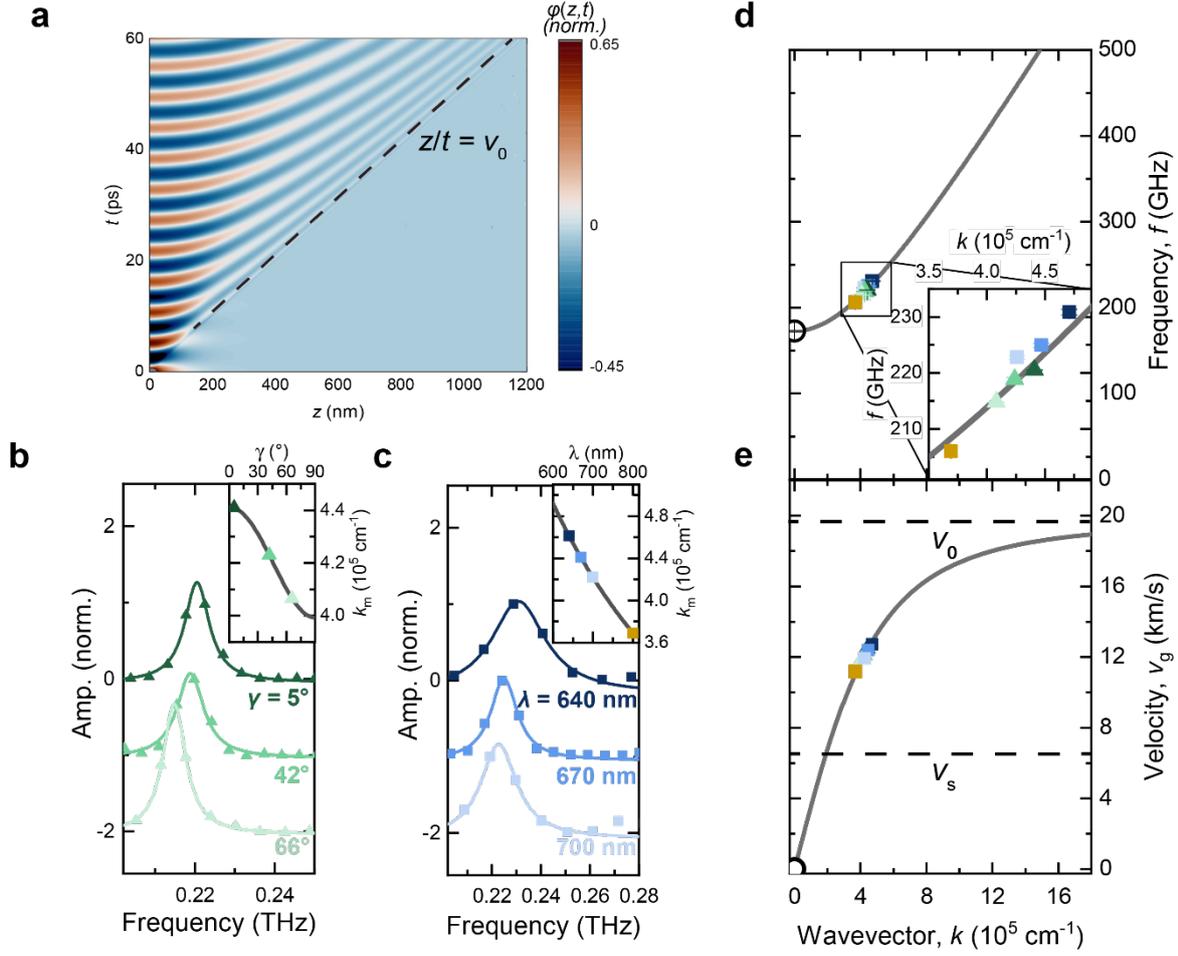

**Figure 4: Revealing spectral components of the broadband antiferromagnetic magnon wavepacket.** (a) Simulation of the time- and position-dependent spin deflection $\varphi(z,t)$ after optical excitation at 3.1 eV with a penetration depth of 50 nm, as determined by Eq. (3). (**b,c**) Fourier spectra of time-resolved measurements of the polarization rotation of a near-infrared probe pulse obtained in the reflection geometry after excitation with pump pulses with a photon energy of 3.1 eV at the temperature $T = 60$K for different probe incidence angles $\gamma$ ($\lambda = 680$ nm) (a) and probe wavelengths $\lambda$ ($\gamma = 5°$) (b). The solid superimposed lines are Lorentzian fits of the Fourier peaks. Insets: The magnon wavenumber $k_m$ to which the probe is sensitive, as a function of the angle $\gamma$ (b) and probe wavelength $\lambda$ (c), with the measured points indicated by coloured markers. (**d,e**) The extracted central magnon frequencies (d) and the calculated group velocity (e) from the data in panel (b) and (c) at their respective calculated wavenumbers plotted with a best fit of the DyFeO$_3$ spin-wave dispersion curve (d) and the group velocity corresponding to the dispersion fit (e). The marker colours and shapes correspond to the measurements in panel (b) and (c).

By optical pumping of above-bandgap electronic transitions, we have explored an efficient and virtually universal route to launch coherent propagating spin waves in insulating antiferromagnets. The strong optical absorption provides an opportunity to spatially confine the light to a subwavelength scale, inaccessible by any other means, e.g. focusing[32-34], enabling the emission of a broadband continuum of short-wavelength antiferromagnetic magnons. The universal mechanism opens up prospects for terahertz coherent AFM magnonics and opto-spintronics[7] providing a long-sought source of coherent high-speed spin-waves. We anticipate even higher propagation velocities to be observed in the broad class of easy-plane antiferromagnets (e.g. hematite[31] and $FeBO_3$), in which the spin-wave gap $\omega_0$ is reduced and the exchange wave regime can be achieved at smaller wavenumbers *k*. The demonstrated approach holds promise for a wide range of fundamental studies exploiting the excitation and propagation of non-linear spin waves such as magnetic solitons[12,35] as well as the investigation of the giant magneto-elastic coupling between antiferromagnetic magnons and acoustic phonons[36] directly in the time-domain.

# Methods

**Sample.**

A single crystal of DyFeO$_3$, 60 μm thick, grown by a floating zone melting technique was used in this work. The sample is cut perpendicularly to the crystallographic *z*-axis.

**Time-resolved experiment.**

The experimental setup is schematically shown in Extended Data Figure 1.

An amplified 1 kHz Ti:Sapphire laser system (Astrella, Coherent, central wavelength 800 nm, pulse energy: 7 mJ, pulse duration: 100 fs) forms the basis of the experimental setup. A large fraction of this output is used to pump a dual optical parametric amplifier (OPA, TOPAS-Twins, Light Conversion). The OPA delivers linearly polarized, 100 fs output pulses, with photon energies in the range 0.45-1 eV ($\lambda = 2.7 – 1.4$ μm). The photon energy $\hbar\omega$ of these output pulses was doubled or tripled using a β-barium borate (BBO) single crystal in order to obtain tunable excitation pulses which cover the photon energies in the optical range of 1.5-3.1 eV (wavelength 400-800 nm). A small portion of the amplifier pulses was sent through a mechanical delay line and used as probe of the spin dynamics in the reflection and transmission geometries.

Pump and probe pulse were focused onto the DyFeO$_3$ sample (pump spot diameter: 300 μm, typical fluence 2 mJ/cm$^2$, probe spot diameter: 80 μm), which was kept in a dry-cycle cryostat (Montana Instruments) that allowed to cool it down to 10 K and vary the temperature with high stability in a wide temperature range (10-250 K). The pump-induced changes in the polarization $\theta_{K,F}$ of the reflected or transmitted probe pulse were measured using an optical polarization bridge (Wollaston prism) and a pair of balanced Si photodetectors.

**Experimental determination of the absorption coefficient.**

The unpolarised absorption spectrum of DyFeO$_3$ was directly obtained with light propagating along the crystal *z*-axis in the spectral region 1-2.2 eV. The resulting absorption is shown in Figure 1b. In addition, we performed spectroscopic ellipsometry measurements using a Woollam M5000 ellipsometer over a wide energy range to obtain the real and imaginary parts of the refractive index. In the photon energy region 2.5-4 eV, where the transmission measurements are not possible for thick samples, we estimated the absorption using the acquired complex refractive index. These values are shown in Figure 1b.

**Data Availability:**

All data presented in this work will be made publicly available with the identifier (DOI) …

**Code Availability:**

The code used to simulate the magnon dynamics is available upon reasonable request.

**Acknowledgements:**

The authors thank V.V. Kruglyak for critical reading of the manuscript and E. Demler for insightful discussions. This work was supported by the EU through the European Research Council, Grant No. 677458 (AlterMateria), the Netherlands Organization for Scientific Research (NWO/OCW) as part of the Frontiers of Nanoscience program (NanoFront) and the VENI-VIDI-VICI program. R.V.M and R.L. acknowledge support from the European Research council, Grant No. 852050 (MAGSHAKE). R.C. acknowledges support by the project Quantox, Grant No. 731473, QuantERA-NET Cofund in Quantum Technologies, implemented within the EU-H2020 Programme. B.A.I acknowledges support from the National Scientific Foundation of Ukraine under Grant. No. 2020.02/0261.

**Author contributions:**

D.A. and A.D.C. conceived the project. J.R.H., D.A. and M.M performed the experiments, analysed the data. B.A.I. developed the general theoretical framework describing the spin wave propagation. R.L and R.V.M. developed the theoretical formalism of the spin wave detection. B.A.I., R.C., R.V.M. and A.V.K. contributed to discussion of the results. A.D.C. supervised the project. The manuscript was written by J.R.H., D.A. and A.D.C., with feedback and input from all co-authors.

**Competing interests:**

The authors declare no competing interests.

# Extended Data

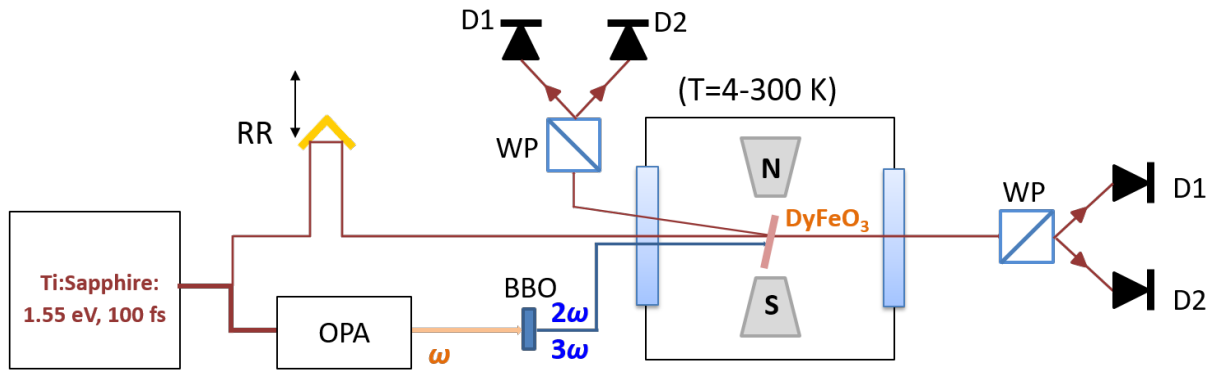

**Extended Data Figure 1: Experimental setup.** RR: gold retroreflector mounted on a motorized precision delay stage, OPA: optical parametric amplifier, BBO: $\beta$-barium borate crystal, WP: Wollaston Prism, D1, D2: pair of balanced silicon photodetectors.

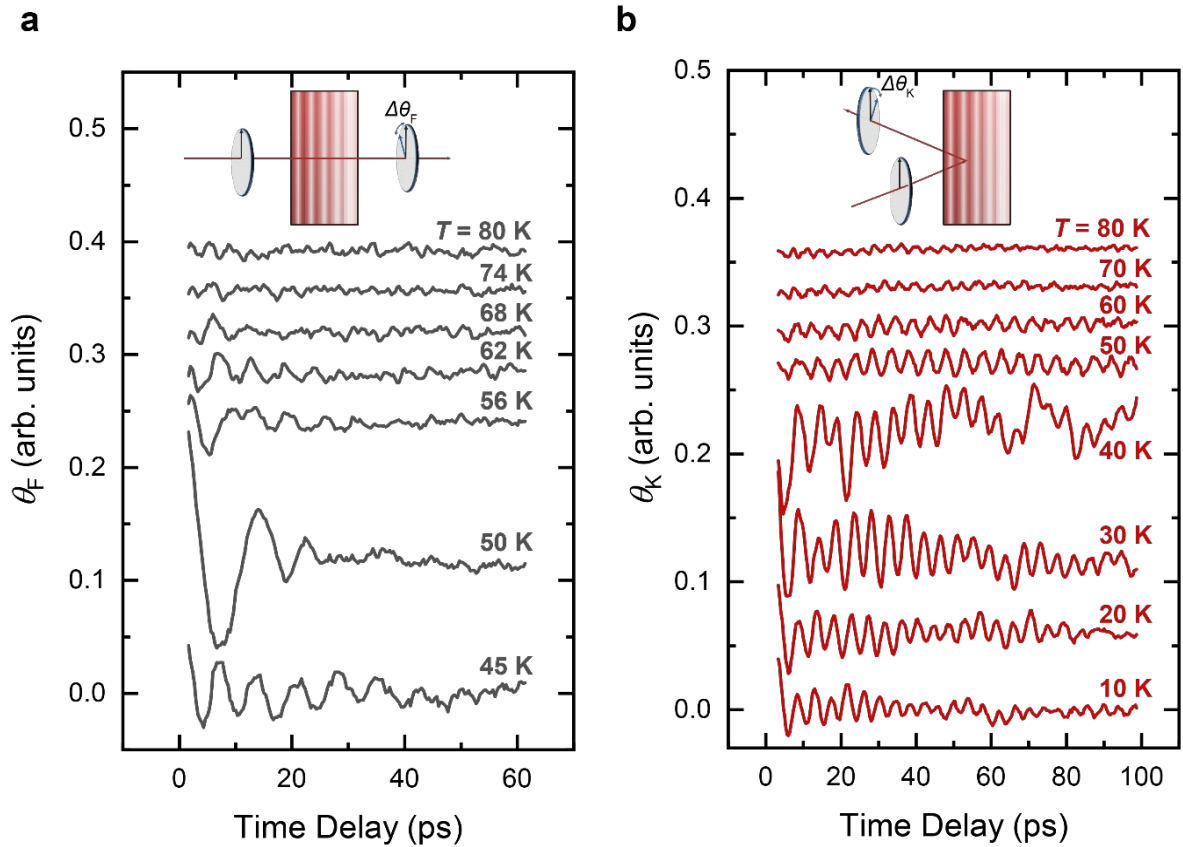

**Extended Data Figure 2: Magnon time traces at different temperatures. a)** Time resolved polarization rotation in the transmission (a) and reflection geometry (b) following excitation at $h\nu = 3.1$ eV at different temperatures. The probe incidence angle is near-normal, with $\lambda = 640$ nm and $\lambda = 700$ nm for the measurements in panel a and b, respectively.

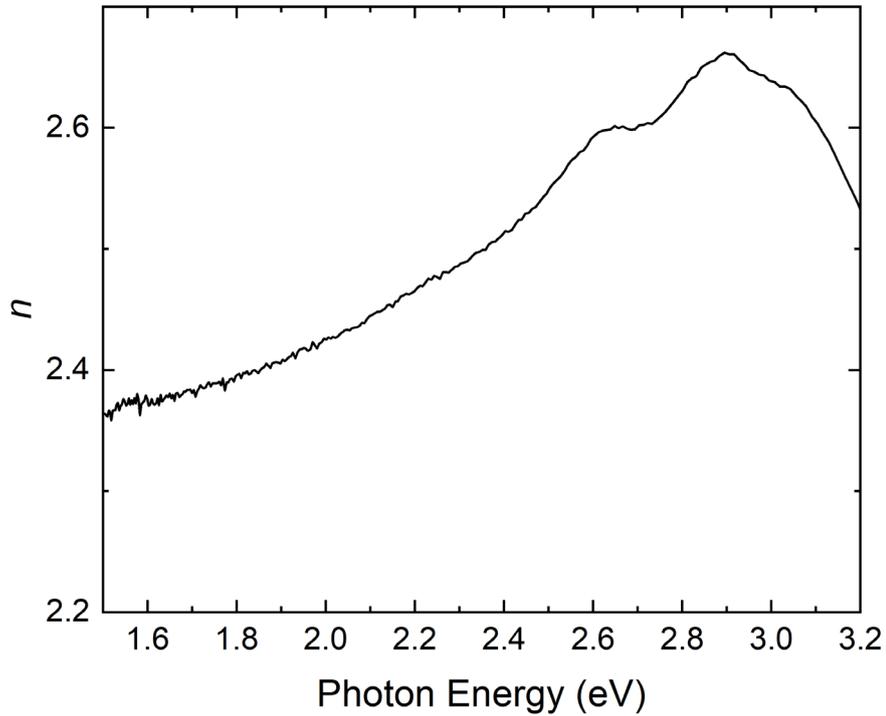

**Extended Data Figure 3: Real part of the refractive index as a function of the pump photon energy.** Real part $n$ of the refractive index, as extracted using spectroscopic ellipsometry measurements.

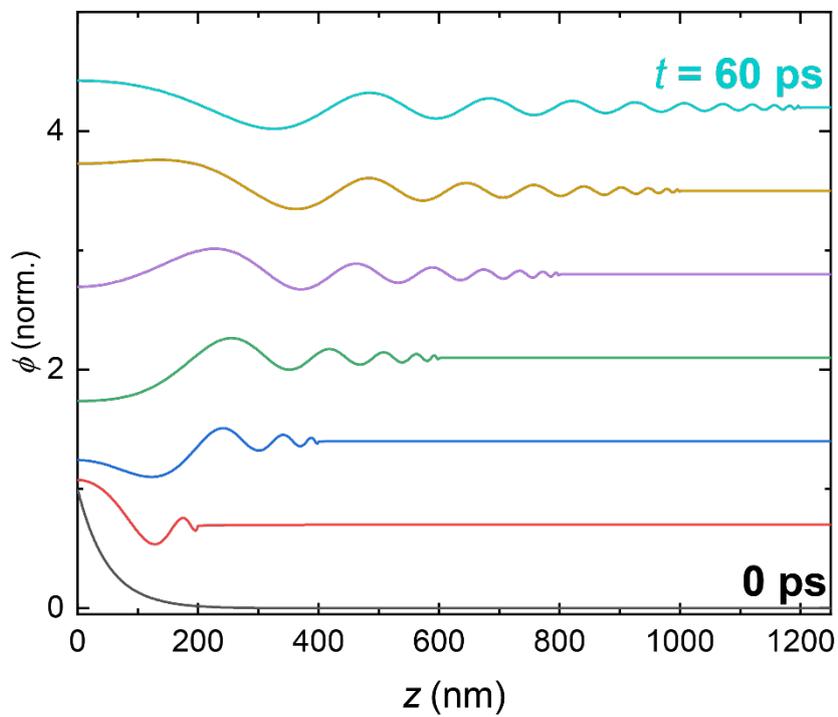

**Extended Data Figure 4: Simulations of the light-induced spin wave dynamics.** Real-space distribution of the magnon spin deflection at different times $t$, after optical excitation at 3.1 eV with a penetration depth of 50 nm, as determined by Eq. (3).

# Supplementary Information

# Coherent spin-wave transport in an antiferromagnet

J.R. Hortensius[1‡*], D. Afanasiev[1‡], M. Matthiesen[1], R. Leenders[2], R. Citro[3], A.V. Kimel[4], R.V. Mikhaylovskiy[2], B.A. Ivanov[5], and A.D. Caviglia[1§]

[1]*Kavli Institute of Nanoscience, Delft University of Technology, P.O. Box 5046, 2600 GA Delft, The Netherlands.*
[2]*Department of Physics, Lancaster University, Bailrigg, Lancaster LA1 4YW, United Kingdom*
[3]*Dipartimento di Fisica "E.R. Caianiello", Università di Salerno and Spin-CNR, I-84084 Fisciano (Sa), Italy*
[4]*Radboud University Nijmegen, Institute for Molecules and Materials, 6525 AJ Nijmegen, The Netherlands.*
[5]*Institute of Magnetism, National Academy of Sciences and Ministry of Education and Science, 03142 Kyiv, Ukraine.*

*Correspondence to: j.r.hortensius@tudelft.nl, dmytro.afanasiev@physik.uni-regensburg.de or a.caviglia@tudelft.nl

[‡]These authors contributed equally to this work

## S1. Theory for magneto-optical detection of propagating spin waves in antiferromagnets

The antiferromagnetic order can be best described by introducing the antiferromagnetic Néel vector $\mathbf{L} = \mathbf{M}_1 - \mathbf{M}_2$, where $\mathbf{M}_{1,2}$ are the magnetizations of the antiferromagnetically coupled sublattices, such that $|\mathbf{M}_1| = |\mathbf{M}_2| = M_0$ and $|\mathbf{L}| \approx 2M_0$. Within the sigma-model approach, the dynamics of the Néel vector is described by the closed equation (sigma-model equation, see, e.g., review article [12] and Supplementary Section S2 for details), whereas the net magnetization $\mathbf{M} = \mathbf{M}_1 + \mathbf{M}_2$ is determined by this vector and its time derivative via the relation:

$$\mathbf{M} = \frac{M_0}{H_{\text{ex}}} \mathbf{H}_D + \frac{1}{2M_0 \omega_{\text{ex}}} \left[ \frac{\partial \mathbf{L}}{\partial t} \times \mathbf{L} \right]. \tag{S1.1}$$

Here $H_{\text{ex}}$ is the effective exchange field ($\omega_{\text{ex}} = 2\gamma H_{\text{ex}}$), with $\gamma = 2\pi \cdot 28$ GHz/T the gyromagnetic ratio, and $\mathbf{H}_D = \frac{H_D}{2M_0}[\mathbf{e}_y \times \mathbf{L}]$ the effective Dzyaloshinskii-Moria field. $\mathbf{e}_y$ is the unit vector along the $y$-axis, which corresponds to the even $C_2$ crystal axis in DyFeO$_3$.

In DyFeO$_3$ the ground state of the Neel vector as well as its dynamics corresponding to the q - AFM mode is restricted to the *(xy)* plane. For such dynamics, the Néel vector can be parametrized by introducing the angle $\varphi_L$ that the vector forms with the $y$-axis, $\mathbf{L} = L_0(\sin\varphi_L, \cos\varphi_L, 0)$. In the ground state of DyFeO$_3$:

$$\varphi_L = \begin{cases} 0, & T < T_M \\ \frac{\pi}{2}, & T > T_M \end{cases} \tag{S1.2}$$

For such planar dynamics, the net magnetization $M_z$ emerges along the $z$-axis, such that:

$$M_z = \frac{M_0}{H_{\text{ex}}} \left( -H_D \sin\varphi_L + \frac{1}{\gamma} \frac{\partial \varphi_L}{\partial t} \right), \tag{S1.3}$$

At this point we introduce a variable $\varphi$ to designate deviations of the vector $\mathbf{L}$ from the equilibrium orientation, such that $\varphi = \varphi_L$ for the collinear AFM state and $\varphi = \pi - \varphi_L$ for the canted AFM state. For the case of small deviations of the Neel vector from the equilibrium, we assume $\varphi \ll 1$ and obtain:

$$M_z = \begin{cases} \dfrac{M_0}{H_{\text{ex}}} \left( -H_D \varphi + \dfrac{1}{\gamma} \dfrac{\partial \varphi}{\partial t} \right), & T < T_M \\ \dfrac{M_0}{H_{\text{ex}}} \left( -H_D - \dfrac{1}{\gamma} \dfrac{\partial \varphi}{\partial t} \right), & T > T_M \end{cases} \tag{S1.4}$$

Optical detection of magnetization dynamics is performed using magneto optical effects, such as the magneto-optical Kerr effect (MOKE) for the case of reflection geometry. The phenomenon originates from a helicity-dependent refractive index in the material with broken time reversal symmetry. The refractive index differs for left-handed and right-handed polarized light, resulting in different reflection coefficients. To calculate the rotation of the plane of polarization after reflection, linearly polarized light is first decomposed into circularly polarized components. For simplicity, it is assumed that the incident light is polarized along the $x$-axis, and the normalized electric field vector $\boldsymbol{e}_i$ in the ($xy$) plane is:

$$\boldsymbol{e}_i = \frac{1}{2}\begin{pmatrix}1\\-i\end{pmatrix} + \frac{1}{2}\begin{pmatrix}1\\+i\end{pmatrix} = \frac{1}{2}\boldsymbol{e}^+ + \frac{1}{2}\boldsymbol{e}^-, \tag{S1.5}$$

where $\boldsymbol{e}^\pm = \begin{pmatrix}1\\\mp i\end{pmatrix}$. Then the reflected field is

$$\boldsymbol{e}_r = \frac{1}{2}r^+\boldsymbol{e}^+ + \frac{1}{2}r^-\boldsymbol{e}^- = \frac{1}{2}\begin{pmatrix}r^+ + r^-\\i(r^- - r^+)\end{pmatrix}, \tag{S1.6}$$

Now the reflectivity is written as the sum of the static reflectivity $r_0$ and the dynamic part of the reflectivity $\Delta r$, which is induced by the spin wave:

$$\begin{aligned}r^+ &= r_0^+ + \Delta r^+\\ r^- &= r_0^- + \Delta r^-\end{aligned} \tag{S1.7}$$

We consider the experimental geometry schematically shown in Fig. S1a, in which a probe pulse enters the material at $z=0$. To find the change in reflectivity, depending on the light helicity, we take an approach similar to the ultrafast detection of acoustic phonons, where phonon-induced strain affects reflectivity [see Ref. [24], Eq. (32)]. In Ref. [24] the change in reflectivity is derived as function of the time and space dependent change in permittivity $\Delta\varepsilon(z,t)$ due to the strain modulation. Here, the same equation is employed to calculate the change of reflectivity induced by magnetization. The equation taken from Ref. [24] reads:

$$r = r_0 + \frac{ik_0^2}{2k}t_0\tilde{t}_0\int_0^\infty dz\, e^{2ikz}\Delta\varepsilon(z,t), \tag{S1.8}$$

where $r_0$ is the reflection coefficient in absence of perturbations in the permittivity, $t_0$ is the transmission coefficient of the light into the medium, and $\tilde{t}_0$ is the transmission coefficient from the medium into free space, $k_0$ is the wave-vector of the light in free space and $k$ is the wave-vector of light in the medium.

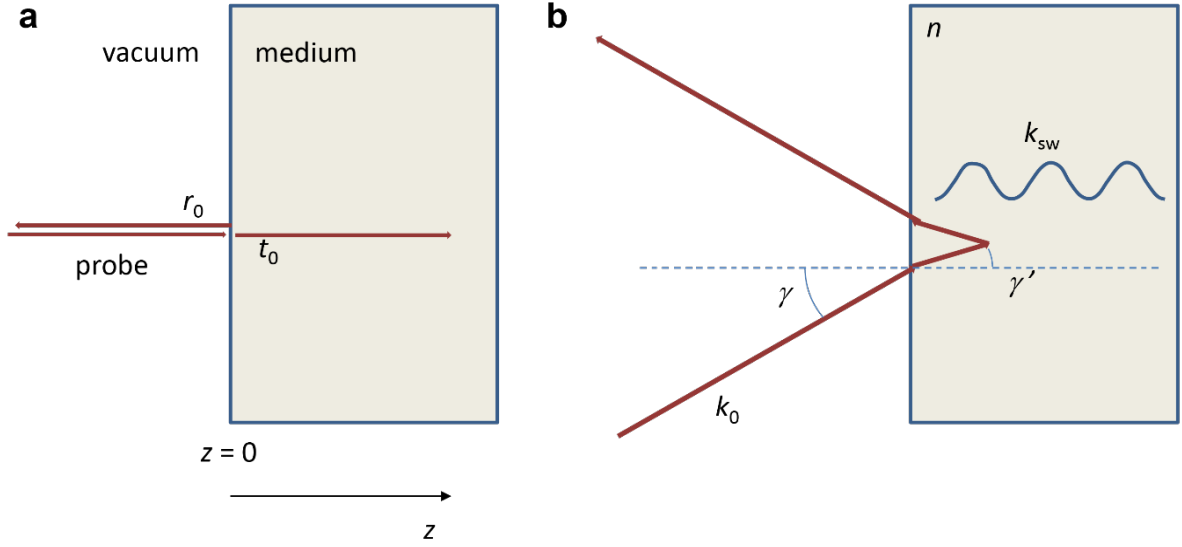

**Figure S1**: (a) Schematic diagram of the experiment considered. (b) Schematic illustration of the detection condition from Eq. (2) main text.

Two electromagnetic eigenmodes exist in a magnetic material with (dynamical) magnetization along the z-axis, which have left-handed and right-handed polarization (±) and experience different refractive indices $n_\pm$[28]. From these effective refractive indices the effective permittivity modulation $\Delta\varepsilon$ can be obtained:

$$n_\pm^2 = \varepsilon \pm ig = \varepsilon + \Delta\varepsilon_\pm, \tag{S1.9}$$

where $g$ is the gyration term. Generally, this gyration term is proportional to the magnetization: $g(M) = aM_z$. From this it is found that:

$$\Delta\varepsilon_\pm(z,t) = \pm iaM_z(z,t) \tag{S1.10}$$

Inserting the expression for $\Delta\varepsilon(z,t)$ in Eq. (S4.8) for right- and left-handed polarization we obtain:

$$\begin{aligned}r^+ &= r_0^+ - t_0^+ \tilde{t}_0^+ \frac{ak_0^2}{2k^+} \int_0^\infty dz' e^{2ik^+ z'} M_z(z',t) \equiv r_0^+ + \Delta r^+ \\ r^- &= r_0^- + t_0^- \tilde{t}_0^- \frac{ak_0^2}{2k^-} \int_0^\infty dz' e^{2ik^- z'} M_z(z',t) \equiv r_0^- + \Delta r^-\end{aligned} \tag{S1.11}$$

For the sake of simplicity, we use the approximation of a pure antiferromagnet, such that the difference in reflection coefficients, transmission coefficients and wave vectors of light with opposite helicity in statics is negligible (we also neglect higher-order effects such as magnetic birefringence), simplifying the expression to:

$$r^+ = r_0 - \Delta r \\ r^- = r_0 + \Delta r \quad, \tag{S1.12}$$

where

$$\Delta r = \frac{a k_0^2}{2k} t_0 \tilde{t}_0 \int_0^\infty dz' e^{2ikz'} M_z(z', t). \tag{S1.13}$$

Now the rotation angle $\theta_K$ is calculated from equation (S1.6), by taking the ratio of the $y$- and $x$-components. Generally, the rotation angles are small such that $\tan(\theta_K) \approx \theta_K$ so that:

$$\theta_K \approx \frac{i(r^- - r^+)}{r^- + r^+} = \frac{i \Delta r}{r_0} \tag{S1.14}$$

The general form of the magnetization emergent in response to propagation of the spin wave packet and driven spin precession can be written as follows:

$$M_z(z, t) = \int_{-\infty}^{\infty} d\omega e^{i\omega t} \left( f(\omega) e^{-i k_{sw}(\omega) z} + p(\omega) e^{-\frac{z}{\delta}} \right) \tag{S1.15}$$

Here, $k_{sw}(\omega)$ is the wave-vector of the spin wave, which is related to $\omega$ through the dispersion relation, $f(\omega)$ and $p(\omega)$ are functions describing the frequency distribution of the propagating spin wave packet and the driven spin precession respectively, and $\delta$ is the optical penetration depth. Substituting Eq. (S1.15) in Eq. (S1.13) and in Eq. (S1.14) afterwards, results in the following expression for the rotation angle:

$$\theta_K = i \frac{t_0 \tilde{t}_0}{r_0} \frac{a k_0^2}{2k} \int_0^\infty dz \int_{-\infty}^{\infty} d\omega e^{i\omega t} \left( f(\omega) e^{i(2k - k_{sw}(\omega))z} + p(\omega) e^{(2ik - 1/\delta)z} \right). \tag{S1.16}$$

We now note that in a general case $k_{sw} = \kappa_{sw} - i\eta_{sw}$, i.e. spin waves decay upon propagation from the sample boundary with decrement $\eta_{sw}$. For the case $\eta_{sw} \neq 0$ the integral over $z$ converges and the result is

$$\theta_K = \frac{t_0 \tilde{t}_0}{r_0} \frac{a k_0^2}{2k} \int_{-\infty}^{\infty} d\omega e^{i\omega t} \left( \frac{f(\omega)}{2k - k_{sw}(\omega)} + \frac{p(\omega)}{2k + \frac{i}{\delta}} \right) \tag{S1.17}$$

with the integral over the frequency representing the inverse Fourier transformation (IFT).

$$\theta_K = -\frac{t_0 \tilde{t}_0}{r_0} \frac{a k_0^2}{2k} \text{IFT} \left[ \frac{f(\omega)}{2k - k_{sw}(\omega)} + \frac{p(\omega)}{2k + \frac{i}{\delta}} \right]. \tag{S1.18}$$

The equations (S1.17) and (S1.18) have a pole for $2k - k_{sw}(\omega)$ implying selective detection of the spin waves with wave vectors satisfying the expression $2k - k_{sw}(\omega) \approx 0$ (assuming $\eta_{sw} \ll \kappa_{sw}$). If one rewrites this expression in terms of the wavelengths $2\lambda_{sw} = \lambda_{opt}$, with $\lambda_{sw}$ the wavelength of the spin wave and $\lambda_{opt}$ the wavelength of the probe pulse in the medium, the well-known Bragg condition is obtained. In the specific case of our experiment, the gradient of the excitation is directed in the $z$-direction, as discussed in Supplementary Section S2, this leads to spin waves with a wavenumber in this particular direction. In the experiment, however, the incoming probe pulse can be directed under a certain angle $\gamma$. Taking refraction into account (see Fig. S1b), which leads to a refracted angle $\gamma'$ for the incoming probe pulse, the detection expression becomes:

$$k_{sw} = 2k_0 n \cos \gamma', \tag{S1.19}$$

which is Eq. (2) in the main text.

The result is the following. Assuming a spin wave packet distribution as discussed in Supplementary Section S2 (see fig S2a), the reflection geometry provides sensitivity to a single component, depending on the probe photon wavenumber. As a result, the oscillation emerging in the time-resolved probe polarization (Fig. S2b) has the frequency of the spin wave at that particular wavenumber.

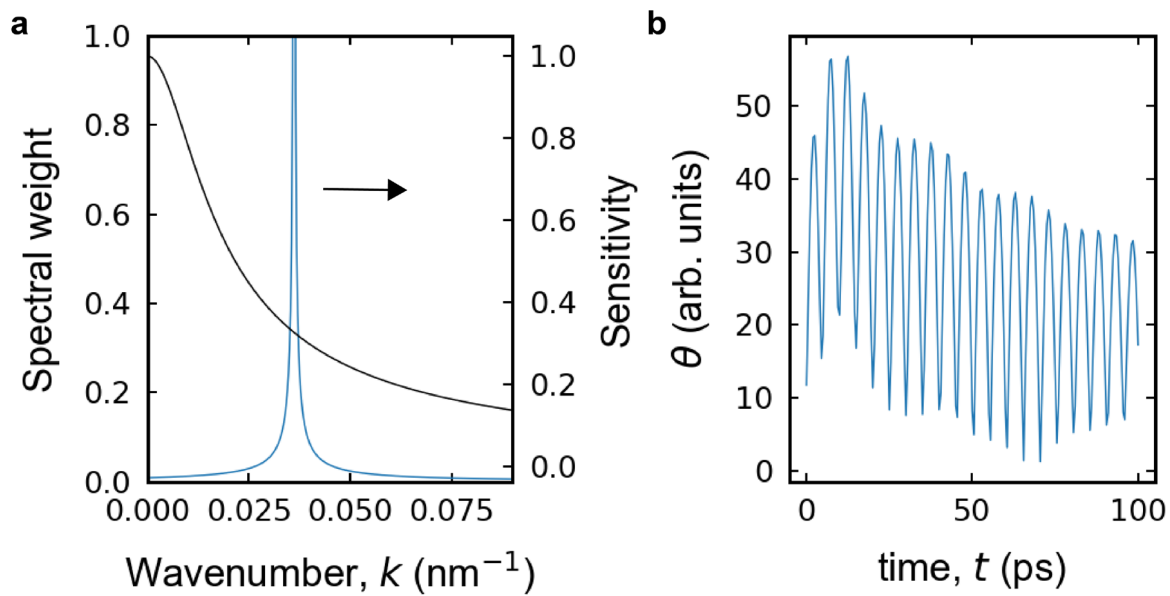

**Figure S2**: a) Wavepacket distribution (black line, left axis) and the sensitivity function as given by the first term of the integrand in Eq. S1.17 (blue line, right axis). b) The resulting time-resolved MOKE signal, given by the wavepacket distribution and the sensitivity function from panel a).

## S2. Theoretical formalism on the generation of the magnon wavepacket

We describe the dynamics of the quasi-antiferromagnetic mode in $DyFeO_3$ using the one-dimensional version of the sigma-model, which for the planar dynamics of the Neel vector can be obtained by the variation of the Lagrangian $L[\varphi]$:[12]

$$L[\varphi] = \int dz \left\{ \frac{A}{2} \left[ \frac{1}{v_0^2} \left(\frac{\partial \varphi}{\partial t}\right)^2 - \left(\frac{\partial \varphi}{\partial z}\right)^2 \right] - w(\varphi) \right\}, \tag{S2.1}$$

where $A$ is the non-uniform exchange constant, $v_0$ is the magnon speed at the linear region of the spectrum, $w(\varphi)$ is the anisotropy energy, and the angle $\varphi = \varphi(z,t)$ describes the deflection of the antiferromagnetic vector $L$ from the equilibrium position (0° and 90° as measured from the $y$-axis in the collinear and canted AFM phase respectively, see Supplementary section S1). Note that the characteristic speed $v_0^2 = \gamma A \omega_{ex}/2M_0$ contains only terms of exchange origin, the uniform exchange parameter $\omega_{ex} = 2\gamma H_{ex}$ and the non-uniform exchange constant $A$, which results in the large value of this speed.

The general equation obtained from (S2.1) is the nonlinear Klein-Gordon equation (note that it transforms to the familiar sine Gordon equation for the variable $2\varphi$ for the simplest form of the anisotropy with only one constant, $w(\varphi) \propto \sin^2 \varphi$). In the linear approximation over the small deviations of $\varphi$ from its equilibrium value it takes the universal form

$$\frac{\partial^2 \varphi}{\partial t^2} - v_0^2 \frac{\partial^2 \varphi}{\partial z^2} + \omega_0^2 \varphi = 0, \tag{S2.2}$$

where $\omega_0$ is the spin-wave gap frequency:

$$\omega_0^2 = \omega_{ex}\omega_a, \qquad \omega_a = \frac{\gamma}{2M_0} \frac{d^2w}{d\varphi^2}\bigg|_{\varphi=0}. \tag{S2.3}$$

The derivative of the anisotropy energy is calculated at the equilibrium value of $\varphi$, see for more details Ref. [22]. Thus, all the quantities can be present through two well-known quantities: $v_0 \approx 20$ km/s [27] and the value of magnon gap $\omega_0$, which is directly measured in our experiment. The characteristic space scale is given by the value $v_0/\omega_0$, of the order of a few nanometres.

Ultrashort pulses of light with a corresponding broadband optical spectrum are routinely being used as an instantaneous excitation to generate high frequency spin dynamics[7]. We start with the assumption that at the time $t = 0$, the spin deflection $\varphi(z,t)$ in the material is given by the profile of the optical excitation, schematically shown in Fig. S3a, as the result of an instantaneous excitation:

$$\varphi(z, t = 0) = \begin{cases} \varphi_0 e^{-\frac{z}{\delta}} & z \geq 0 \\ 0 & z < 0 \end{cases}. \tag{S2.4}$$

Here, $z=0$ forms the interface between the material and vacuum, $\delta$ is the penetration depth of the excitation pulse and $\varphi_0$ the amplitude of the initial spin deflection, proportional to the pump fluence and inversely proportional to $\delta$ ($\varphi_0 \sim I/\delta$), as to keep the total absorbed energy for pulses with different penetration depths equal to the intensity $I$ of the pulse.

In order to account for the boundary condition given by the surface of the sample, we look at the energy flow $j_E$; from the Lagrangian (Eq. S2.1) it follows:

$$j_E = -\frac{\partial L}{\partial(\partial\varphi/\partial z)} \frac{\partial \varphi}{\partial t} = -A \frac{\partial \varphi}{\partial z} \frac{\partial \varphi}{\partial t}. \tag{S2.5}$$

The energy flow should vanish at the surface ($z = 0$) at all times $t$. This gives the boundary condition $\frac{\partial \varphi}{\partial z}\big|_{z=0} = 0$. The simplest way to find a solution obeying this boundary condition is to expand the problem symmetrically to $z < 0$, such that the solution of the symmetrical problem is $\tilde{\varphi}(z,t) = \tilde{\varphi}(-z,t)$ and can be found with the initial conditions:

$$\tilde{\varphi}(z, t = 0) = \varphi_0 e^{-\frac{|z|}{\delta}} \tag{S2.6a}$$

and

$$\frac{\partial \varphi}{\partial z}(z, t = 0) = \frac{\varphi_0}{\delta} \begin{cases} -e^{-\frac{z}{\delta}} & z \geq 0 \\ 0 & z = 0 \\ e^{\frac{z}{\delta}} & z < 0 \end{cases} \tag{S2.6b}$$

Using the dispersion relation of the material, which is obtained after solving Eq. S2.2, we obtain that for a given spin wave component $\psi_k$:

$$\psi_k(z,t) = A e^{ikz - i\omega_k t} + B e^{ikz + i\omega_k t}, \quad \omega_k = \sqrt{\omega_0^2 + (v_0 k)^2} \tag{S2.7}$$

Here, as before, $\omega_0 = \sqrt{\omega_{ex}\omega_a}$ the spin wave gap and $v_0$ the characteristic speed. Having in mind the symmetry of the wanted solution, the only solution which symmetric over inversion of the magnon eigenmodes, is:

$$\psi_k(z,t) = C_k \cos(kz)\cos(\omega_k t) \tag{S2.8}$$

From Eq. (S2.6a) we can present the spin deflection $\tilde{\varphi}(z, t = 0)$ in a material slab of thickness $d$ using the Fourier expansion as:

$$\tilde{\varphi}(z, t = 0) = \varphi_0 \sum_k a_k e^{ikz},$$

$$a_k = \frac{1}{d} \frac{2\delta}{1 + (\delta k)^2}$$

(S2.9)

From this expression we find that the initial exponential distribution in spin-deflection in real-space corresponds to a broadband magnon wavepacket, as shown in Fig. S3b.

Then in the continuous limit ($d \to \infty$), combining Eq. (S2.4) and Eq. (S2.9) we can easily obtain the final expression, given in the manuscript:

$$\varphi(z, t) = \tilde{\varphi}(z \geq 0, t) = \frac{2\varphi_0}{\pi} \int_{-\infty}^{\infty} dk \left[ \frac{\delta}{1 + (k\delta)^2} \cos(kz) \cos(\omega_k t) \right]. \quad (z \geq 0), \quad \varphi_0 \sim \frac{I_0}{\delta} \quad (S2.10)$$

Eq. S2.10 describes the spin deflection as a function of space ($z$) and time ($t$) and therefore the dynamics of the broadband magnon wavepacket which will start propagating into the sample. This is shown in Fig. 4a in the main text.

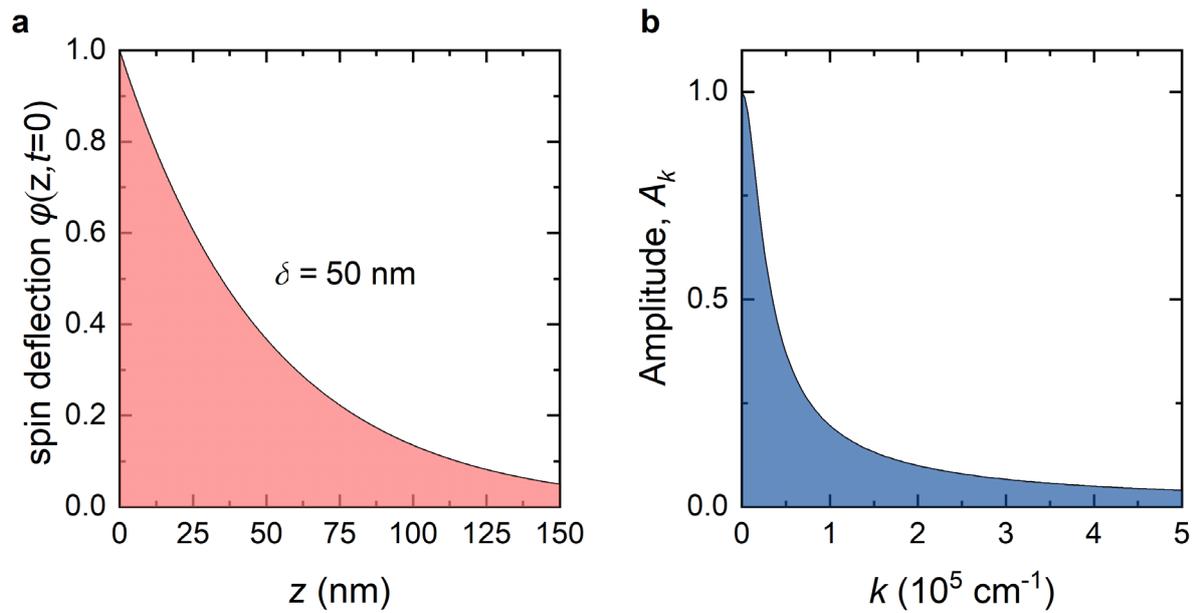

**Figure S3: Real-space wavevector spin deflection distribution.** Real-space distribution of the spin deflection $\varphi(z,t)$ (a) and the corresponding wavevector distribution (b).

## S3. Extracting the spin wave lifetime

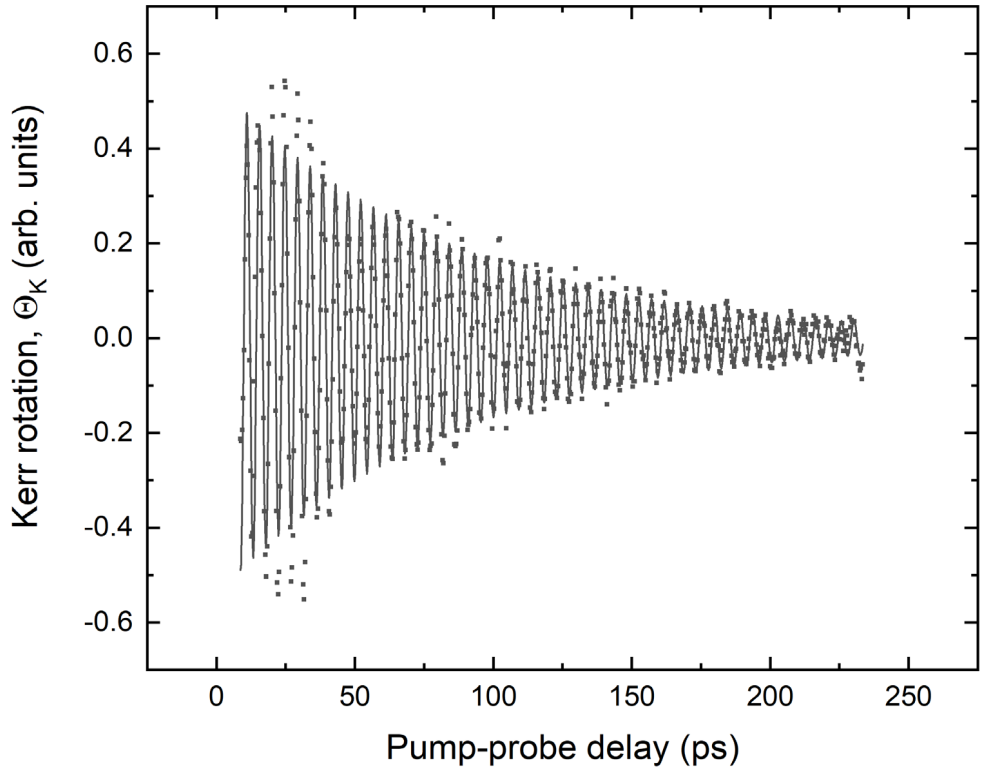

**Figure S4:** Time-resolved polarization rotation originating from a propagating magnon, as obtained in the reflection geometry. The solid line represents a best fit of a damped sine, giving a lifetime of ~85 ps.

The propagation distance can be calculated using the extracted lifetime (see Fig. S4) $\tau = 85$ ps and with the largest estimated group velocities $v_g$ of the measured magnons of about 13 km/s. This gives a propagation distance $l_c = v_g \tau = 1.1$ μm.

## S4. Excitation of a propagating wavepacket of coherent acoustic phonons

Following the excitation with pump pulses at a photon energy of 3.1 eV, the time-resolved polarization rotation signal $\theta_K$ reveals oscillatory dynamics at two central frequencies (see Fig. S5a). As argued in the main text, the high-frequency component corresponds to a finite-$k$ magnon mode. We argue below that the slow-varying component is the result of interference caused by a broadband wavepacket of propagating acoustic phonons as in details described in Ref. 27. This slow-varying oscillating component has been subtracted for the data presented in main text Fig 2b. In this supplementary section we discuss the physical origin of the opto-acoustic signal.

The generation and detection of ultrafast light-induced coherent acoustic phonons in solids is a well-established research field known as picosecond acoustics[37]. Typically, the generation is based on a conversion of the energy of the strongly absorbed ultrashort pump pulse into photo-induced stress in vicinity of the material surface[37]. The photo-induced stress promotes a front of acoustic phonons propagating into the material with the speed of sound $v_s$. In a material that is optically transparent to the probe pulse, the propagation of the acoustic wavepacket can be probed in real time, as it causes interference between light pulses reflected at the crystal surface and at the strain pulse position (see Fig. S5b). This results in oscillations in the probe pulse polarization rotation with a frequency $f_a$ given by [38,39]:

$$f_a = 2nv_s \frac{\cos(\varphi)}{\lambda} . \quad (S4.1)$$

Here $n$ is the refractive index of the material at the probe wavelength $\lambda$ and $\varphi$ the angle with respect to the sample normal.

To validate our conjecture, the measured oscillations can be used to extract the speed of sound in DyFeO$_3$ using Eq. (S4.1) and compared to literature values. Extracting the central frequency of the slow oscillation ($f_a$ = 42.6 GHz), and using the refractive index $n$ = 2.32 for DyFeO$_3$ at the probe wavelength of 700 nm and near-normal incidence ($\varphi \approx 0$), we find that $v_s$ = 6.4·10$^3$ km/s. This agrees well with literature values of sound velocities in the orthoferrites, which are all about 6.0 km/s at cryogenic temperatures[40-42]. The opto-acoustic conversion process underlying the detection is strongly peaked at a photon wavenumber $k_a$ determined by the probe photon wavenumber, equal in magnitude to the value in the magnon detection mechanism[43].

We therefore conclude that optical excitation of the strongly-absorbing charge-transfer resonances excites both a propagating broadband magnon wavepacket and a propagating broadband acoustic phonon wavepacket.

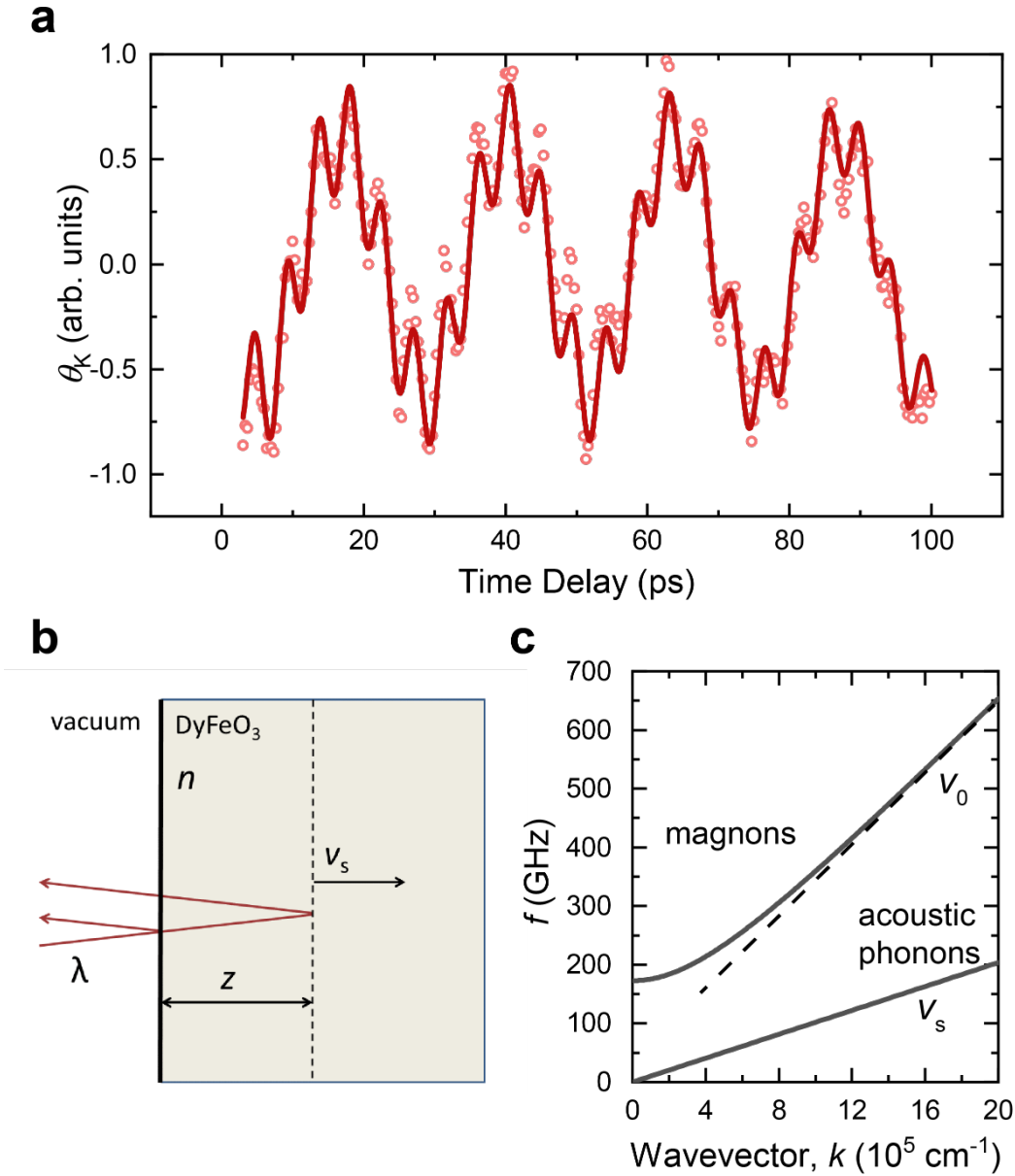

**Figure S5: Observation of a wavefront of propagating acoustic phonons. a**) Time resolved polarization rotation in the reflection geometry following excitation at $h\nu = 3.1$ eV. The solid line represents a best fit using a double-sine function. The slow oscillations are the result of a propagating acoustic wavefront. The higher frequency oscillation corresponds to a finite $k$ magnons mode. **b**) Schematic picture of an acoustic strain wave propagating with velocity $v_s$, giving interference between the light pulse with wavelength $\lambda$ reflected from the vacuum-DyFeO$_3$ interface and scattered at the strain wave. **c**) Dispersion relation for magnons and (longitudinal) acoustic phonons in DyFeO$_3$. The slope of the dispersion defines the (limiting) propagation speed of the waves.